\documentclass[twocolumn,amsmath,amssymb,aps,prl,floatfix]{revtex4-2}

\usepackage{graphicx}% Include figure files
\usepackage{dcolumn}% Align table columns on decimal point
\usepackage{bm}% bold math
\usepackage{siunitx}
\usepackage{nicefrac}
\usepackage[super]{nth}
\usepackage{color}

\newcommand{\ket}[1]{\mathinner{|{#1}\rangle}}
\newcommand{\bra}[1]{\mathinner{\langle{#1}|}}

\newcommand{\mycolumnwidth}{\linewidth}
\newcommand{\rph}{A} 
\newcommand{\bph}{393} 

\begin{document}

\preprint{APS/123-QED}

\title{Full Bell-basis measurement of an atom-photon 2-qubit state and its application for quantum networks}

\author{Elena Arensk\"otter}
\thanks{} 
\affiliation{Experimentalphysik, Universit\"at des Saarlandes, 66123 Saarbr\"ucken, Germany}

\author{Stephan Kucera}
\thanks{}
\affiliation{Experimentalphysik, Universit\"at des Saarlandes, 66123 Saarbr\"ucken, Germany}

\author{Omar Elshehy}
\thanks{}
\affiliation{Experimentalphysik, Universit\"at des Saarlandes, 66123 Saarbr\"ucken, Germany}

\author{Max Bergerhoff}
\thanks{}
\affiliation{Experimentalphysik, Universit\"at des Saarlandes, 66123 Saarbr\"ucken, Germany}

\author{Matthias Kreis}
\thanks{}
\affiliation{Experimentalphysik, Universit\"at des Saarlandes, 66123 Saarbr\"ucken, Germany}

\author{Léandre Brunel}
\thanks{Current address: Department of Physics, University of Virginia, Charlottesville, Virginia 22903, USA}
\affiliation{Experimentalphysik, Universit\"at des Saarlandes, 66123 Saarbr\"ucken, Germany}

\author{J\"urgen Eschner}
\thanks{\href{mailto:juergen.eschner@physik.uni-saarland.de}{juergen.eschner@physik.uni-saarland.de}}
\affiliation{Experimentalphysik, Universit\"at des Saarlandes, 66123 Saarbr\"ucken, Germany}

\date{\today}% It is always \today, today,

\begin{abstract}
The efficiency of a Bell-state measurement on photon pairs is bound to 50\,\% due to the number of Bell states that can be distinguished using linear optics. Here we present the implementation of a protocol that allows us to distinguish all four Bell states by the use of a single-ion quantum memory and heralded absorption as state-selective measurement. The protocol is implemented in two steps. First we demonstrate the state-preserving mapping of a photonic qubit onto the quantum memory, verified by the preservation of entanglement in the process. Then we demonstrate the full Bell state projection between a memory qubit and an incoming photonic qubit, by applying it for atom-to-photon quantum state teleportation.
\end{abstract}

\maketitle
Photonic Bell state projection is a fundamental tool in quantum networks based on single photons. It enables, for example, the extension of a quantum link with quantum repeater operations \cite{Briegel1998} and the transmission of quantum information via quantum teleportation \cite{Bouwmeester1997}. A generic application, that also provides the context to this article, is the generation of remote memory entanglement via photonic Bell-state projection at a central station \cite{Moehring2007, Leent2022, Bernien2013}. Bell state discrimination on photon pairs is fundamentally limited to 50\,\% \cite{Braunstein1995, Michler1996} when only linear optical elements are used. Overcoming this limitation becomes possible using hyper-entanglement \cite{Kwiat1998, Schuck2006, Barbieri2007, Li2017}, nonlinear optical elements \cite{Kim2001}, or auxiliary photons \cite{Bayerbach2023}. An alternative implementation for entangling remote memories, that allows one to distinguish between all four Bell states, was demonstrated in \cite{Welte2021} with trapped Rb atoms. This approach does not use a central station but is mediated by successive interaction of two photons with both memory qubits. The gain in fundamental efficiency is, however, partially offset by the condition that both photons need to travel the full distance. Another possibility is the use of memories at the central station, to which the photonic qubits are transferred before a projective measurement. Using quantum gates on single ions in the same trap, deterministic projection onto all four two-atom Bell states has been demonstrated \cite{Riebe2004, Barrett2004}. Combining this with photon-to-atom state mapping \cite{Kurz2014} would allow full Bell state projection on two incoming photons. 
 
In this manuscript we present a related protocol that allows us to distinguish all Bell states, using only a single memory qubit in a $^{40}$Ca$^+$ ion. In this approach, a first incoming photon is mapped onto the memory qubit. A full Bell state projection is then performed between the memory qubit and the second incoming photon by heralded absorption \cite{Kurz2016}, using projection of the herald and of the memory qubit as Bell state measurement.

We demonstrate the elements of the protocol in two separate steps: the first is the faithful mapping of an incoming single photon onto the memory qubit. For this we apply the heralded absorption interface that was introduced in \cite{Kurz2014} and verified with laser photons. We use the interface to store in the single ion a single photon, heralded by and entangled with its partner from an SPDC photon pair source \cite{Arenskoetter2023}. We characterize the mapping by the fidelity with which the original photon-photon entanglement is preserved in the resulting atom-photon state. The second step is the projection of a 2-qubit state between the memory and an incoming photon onto the four Bell states. We demonstrate and verify this full Bell measurement, which again is facilitated by heralded absorption, by its direct application for quantum state teleportation from the prepared memory qubit onto the partner photon of the absorbed one. The observation of the expected, Bell-state-specific, unitary rotation benchmarks the fidelity of the teleportation and hence the Bell measurement. While the overall efficiency in our proof-of-principle experiment is severely limited, application of the protocol in combination with resonator-based interfaces \cite{Schupp2021, Brekenfeld2020} holds the perspective of overcoming the limitations of linear optical projections.

\section{Experimental setup}

\begin{figure}[t]
	\centering
	\includegraphics [width=\columnwidth]{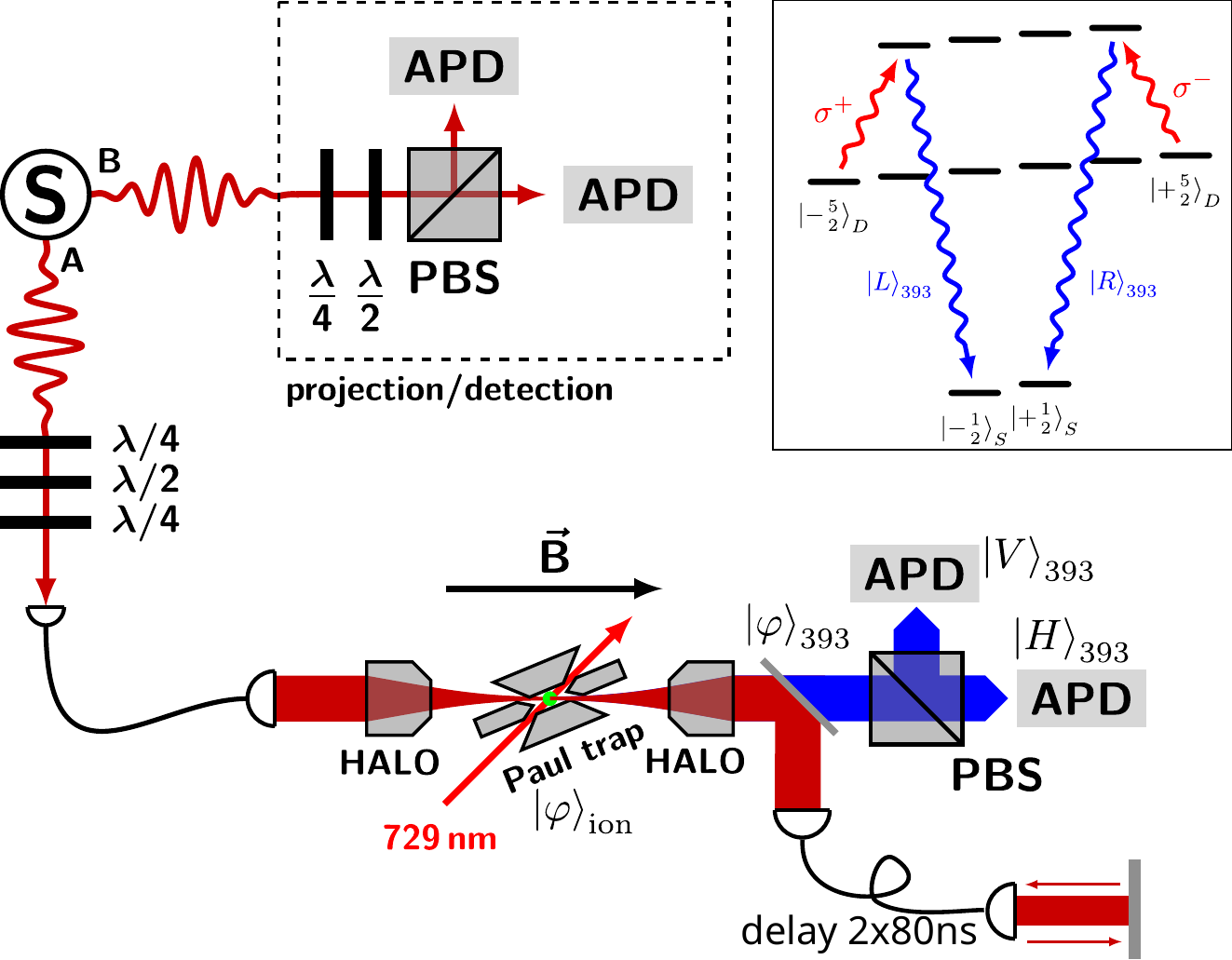}
	\caption{\label{fig.setup} Schematic of the experiment. The photon-pair source S generates polarization-entangled photon pairs. Photons in output A are directed to the atomic setup, where a single trapped $^{40}$Ca$^+$ ion acts as a quantum memory. The photons pass a set of waveplates to compensate for polarization rotations in the fiber connecting the photons to the atomic setup. The partner photons in output B are sent to a tomography setup for analyzing their polarization state. The inset shows the relevant energy levels and transitions of $^{40}$Ca$^+$. Further details are provided in the text. } 
\end{figure}

Figure \ref{fig.setup} shows the experimental setup and the relevant energy levels and transitions of $^{40}$Ca$^+$. A single $^{40}$Ca$^+$ ion is trapped in a linear Paul trap with single-photon optical access by two in-vacuum high-numerical-aperture laser objectives (HALOs, $\text{NA} = 0.4$). One of the objectives is used to focus \SI{854}{\nano\meter} photons onto the ion, in order to excite it from its initial state in the D$_{5/2}$ manifold to the P$_{3/2}$ manifold. The second HALO is used to collect the emitted (Raman-scattered) \SI{393}{\nano\meter} photons which are subsequently polarisation-projected with a Wollaston prism (PBS) and detected with free-space coupled avalanche photo diodes (APDs) in both outputs. This HALO is also used to couple the non-absorbed \SI{854}{\nano\meter} photons into a \SI{16}{\meter} long single-mode fiber with attached retroreflector. The fiber is used to create $\SI{160}{\nano\second}$ delay and an inversion of the direction relative to the quantization axis, which is defined by a magnetic field of \SI{2.855}{G}.
 
\begin{figure}[tbh]
	\centering
	\includegraphics[width=\mycolumnwidth]{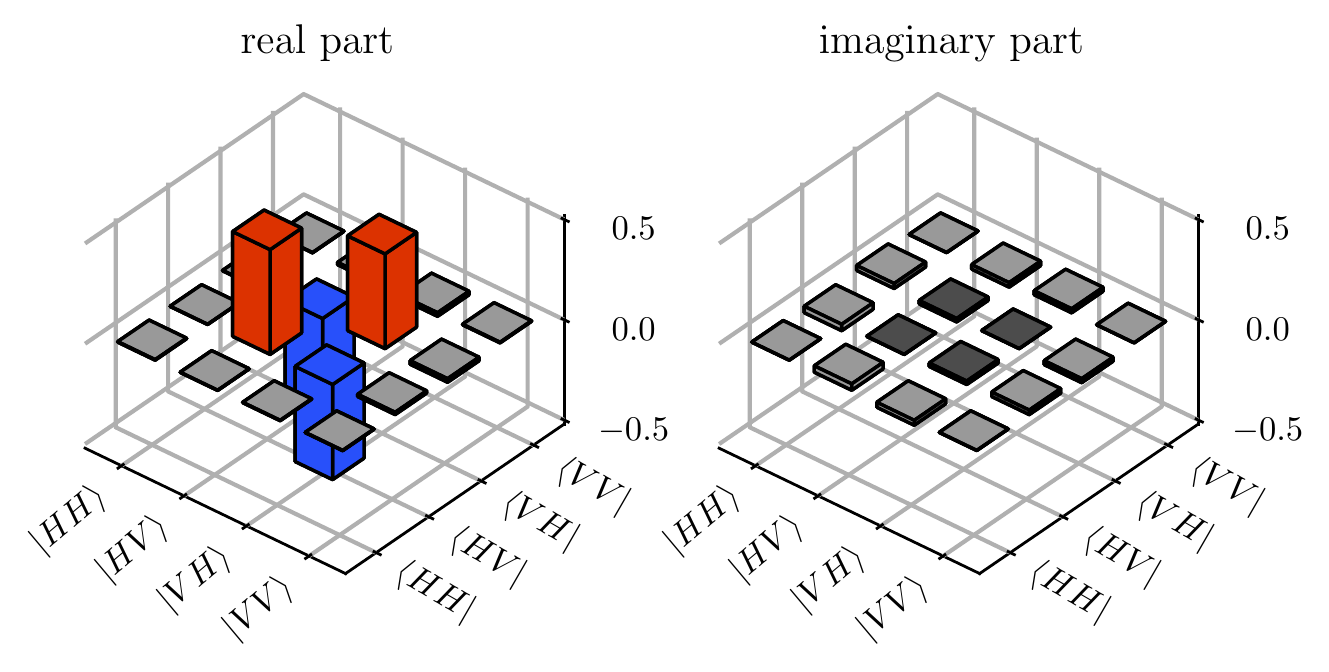}
	\caption{\label{fig.PhRho} Real and imaginary part of the photon-photon density matrix, reconstructed with maximum-likelihood quantum state tomography \cite{James2001}.}
\end{figure}

The photon pair source, labelled S, produces frequency-stable, narrow-band, polarization-entangled photon pairs at \SI{854}{\nano\meter} in a cavity enhanced spontaneous-parametric-down-conversion process (SPDC) with interferometric configuration. More detailed information is provided in \cite{Arenskoetter2023}. The photon in output A has a linewidth of $\SI{12.29}{\mega\hertz}$ and is tuned to be resonant with the D$_{5/2}$-P$_{3/2}$ transition of the $^{40}$Ca$^+$ ion ($\SI{22}{\mega\hertz}$-linewidth).  Output B is detuned by \SI{+480}{\mega\hertz}. The result of a tomographic measurement \cite{James2001} on the photon-pair state $\rho$ in the fiber-coupled outputs A and B is shown in figure \ref{fig.PhRho}.
A fidelity of $\bra{\Psi^-}\rho\ket{\Psi^-} = \SI{91.64(2)}{\percent}$ (\SI{97.16(2)}{\percent} with background correction) with the antisymmetric Bell state, and a purity of $\text{tr}(\rho^2) = \SI{84.97(4)}{\percent}$ (\SI{95.21(3)}{\percent} with background correction) are measured at our operating conditions (15~mW of pump light at 427~nm). Background originates from lost-partner events and detector dark counts; see \cite{Arenskoetter2023} for more information. A fiber-coupled pair rate of \SI{2.69e+05}{\per\second} in the output fibers A and B is inferred from the correlation measurements.

\section{Step\,1: Qubit mapping}

We now describe the mapping of a single-photon qubit state to a qubit state in a single trapped ion and characterize it by its process fidelity, as well as by quantum state tomography. The mapping protocol is based on \cite{Mueller2014, Kurz2014} and adapted to the experimental setup shown in figure \ref{fig.setup}. We use the abbreviations $\ket{\pm\nicefrac{1}{2}}_S = \ket{\text{S}_{1/2},\ m = \pm \nicefrac{1}{2}}$ and $\ket{\pm\nicefrac{5}{2}}_D = \ket{\text{D}_{5/2},\ m = \pm \nicefrac{5}{2}}$ for the atomic states and $\ket{R}$/$\ket{L}$ ($\ket{H}$/$\ket{V}$) for the single-photon states in the circular (linear) polarization modes. The protocol starts by preparing a symmetric coherent superposition in $\text{D}_{5/2}$, $\ket{\varphi}_D = \frac{1}{\sqrt{2}} \, \left( \ket{-\nicefrac{5}{2}}_D + e^{i\, \phi} \ket{+\nicefrac{5}{2}}_D \right)$, then the photon-pair generation is switched on with an acousto-optic modulator that controls the SPDC pump laser. Photons from arm A are focused onto to the ion, co-propagating with the direction of the magnetic field. The absorption of an \SI{854}{\nano\meter} (red) photon in the polarization qubit state
\begin{equation}
\ket{\varphi}_{\rph} = a \ket{R}_{\rph} + b \ket{L}_{\rph}~,
\end{equation}
with amplitudes $a$ and $b$, releases with high probability (\SI{93.5}{\percent}) a \SI{393}{\nano\meter} (blue) Raman photon in the state $\ket{\varphi}_{\bph}$. The blue photon is detected and polarization-analysed with 1.64\,\% probability. If the red photon is not absorbed in the first passage, it is collected and sent back to the ion with a time delay of $\SI{160}{\nano\second}$, now counter-propagating to the magnetic field. Absorption in the first and second passage are distinguished by the delay %and the strong time correlation 
of the detected \SI{393}{\nano\meter} photon with respect to the partner photon in arm B. 

The Raman scattering process is described by the operators \cite{Mueller2014}
\begin{equation}
\begin{split}
\hat{\text{R}}_{\rph,D}^{\text{\nth{1}}} = \ \ 
&\ket{L}_{\bph} \ket{-\nicefrac{1}{2}}_S  \bra{R}_{\rph} \bra{-\nicefrac{5}{2}}_D\\
+&\ \ \ket{R}_{\bph} \ket{+\nicefrac{1}{2}}_S \bra{L}_{\rph} \bra{+\nicefrac{5}{2}}_D 
\end{split}
\label{RamanOpFirst}
\end{equation}
\begin{equation}
\begin{split}
\hat{\text{R}}_{\rph,D}^{\text{\nth{2}}} = \ \ 
&\ket{L}_{\bph} \ket{-\nicefrac{1}{2}}_S  \bra{L}_{\rph} \bra{-\nicefrac{5}{2}}_D\\
+&\ \ \ket{R}_{\bph} \ket{+\nicefrac{1}{2}}_S \bra{R}_{\rph} \bra{+\nicefrac{5}{2}}_D 
\end{split}
\label{RamanOpSecond}
\end{equation}
for the first and second passage, respectively. Detecting  $\ket{\varphi}_{\bph}$ in either of the states  
\begin{equation}
\begin{split}
\ket{H}_{393} = \frac{ \ket{R}_{393} \text{+} \ket{L}_{393}} {\sqrt{2}}
\\
\ket{V}_{393} = \frac{ \ket{R}_{393} \text{--} \ket{L}_{393}}{i\,\sqrt{2}} 
\label{protocolLinPol}
\end{split}
\end{equation}
completes the state-mapping process from $\ket{\varphi}_{\rph}$ to the atomic ground state qubit, which is now described by
\begin{equation}
\ket{\varphi}_S^{\text{\nth{1}}} = \frac{1}{\sqrt{2}} \, \left( a \ket{-\nicefrac{1}{2}}_S \pm b \ket{+\nicefrac{1}{2}}_S \right)
\end{equation}
\begin{equation}
\ket{\varphi}_S^{\text{\nth{2}}} = \frac{1}{\sqrt{2}} \, \left( b \ket{-\nicefrac{1}{2}}_S \pm a \ket{+\nicefrac{1}{2}}_S \right)
\end{equation}
for the first and second passage; the $\pm$ sign corresponds to the $\ket{H}_{393}$ and $\ket{V}_{393}$ projection results of the \SI{393}{\nano\meter} photon, which also heralds the absorption and thereby allows us to filter out successful mapping processes. 
The fidelity of the mapping, expressed in terms of process fidelity \cite{Chuang1997} was determined to be $\chi_{11} = \SI{96.2}{\percent}$ by an independent measurement using single photons; see the Supplement for more information. Magnetic field fluctuations are the main limitation to the fidelity. To counteract phase fluctuations in the atomic qubit due to magnetic stray fields, a spin-echo sequence synchronised with the Larmor precession is applied in the following measurements (see Methods).  

Another means of characterizing the photon-to-atom state mapping is provided by measuring the transfer of polarization entanglement from the two-photon state to the atom-photon state.
Using all heralded absorption events, full quantum state tomography is performed on the two-qubit state of atom in S$_{1/2}$ and 393-nm photon, similar to \cite{Bock2018}. It reveals the reconstructed density matrices displayed in figure \ref{fig.AtPhRho}. Details on the data and the reconstruction procedure are available in the Supplement. The overlap fidelities and purities of the reconstructed states with the ideal, maximally entangled states ($\ket{\Psi^-}$ in the $1^\textrm{st}$ passage and $\ket{\Phi^-}$ in the $2^\textrm{nd}$ passage) are summarized in table \ref{tab:entanglementResults}. The values in brackets are corrected for background and binning. The most relevant background contribution is caused by accidental coincidences, inherent to an SPDC pair source. The binning correction is described in the Methods section. The two figures of merit serve different purposes: a practical system, set to operate at a fixed pair rate, would most honestly be characterized by the uncorrected fidelities that quantify, for example, the usable entanglement. In contrast, characterization of the implementation and operation of the protocol would exclude such background, as well as binning and detector noise, such that the corrected fidelities apply. 
\begin{figure}
	\centering
	(a)\includegraphics[width = \mycolumnwidth]{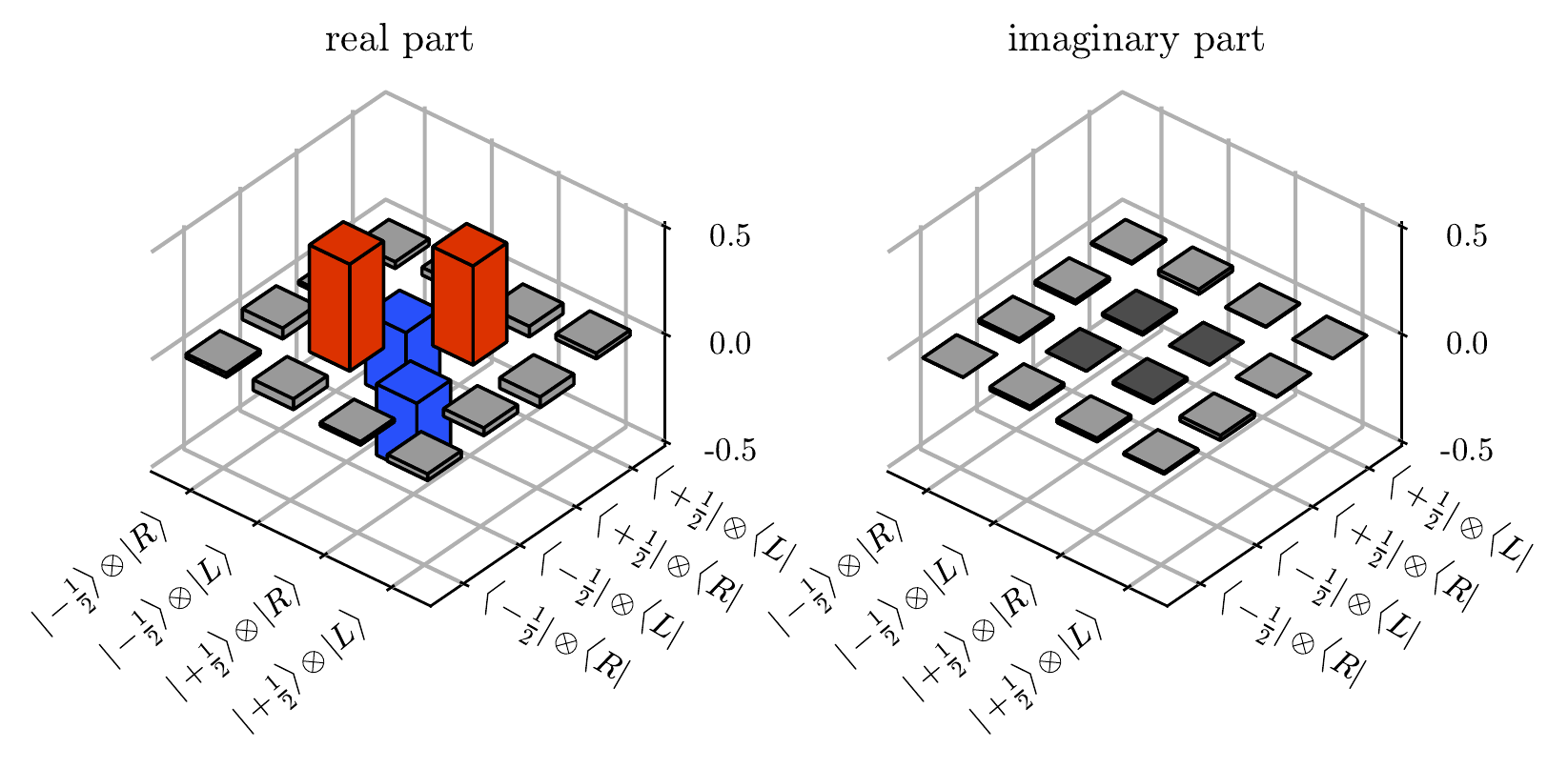}
	(b)\includegraphics[width = \mycolumnwidth]{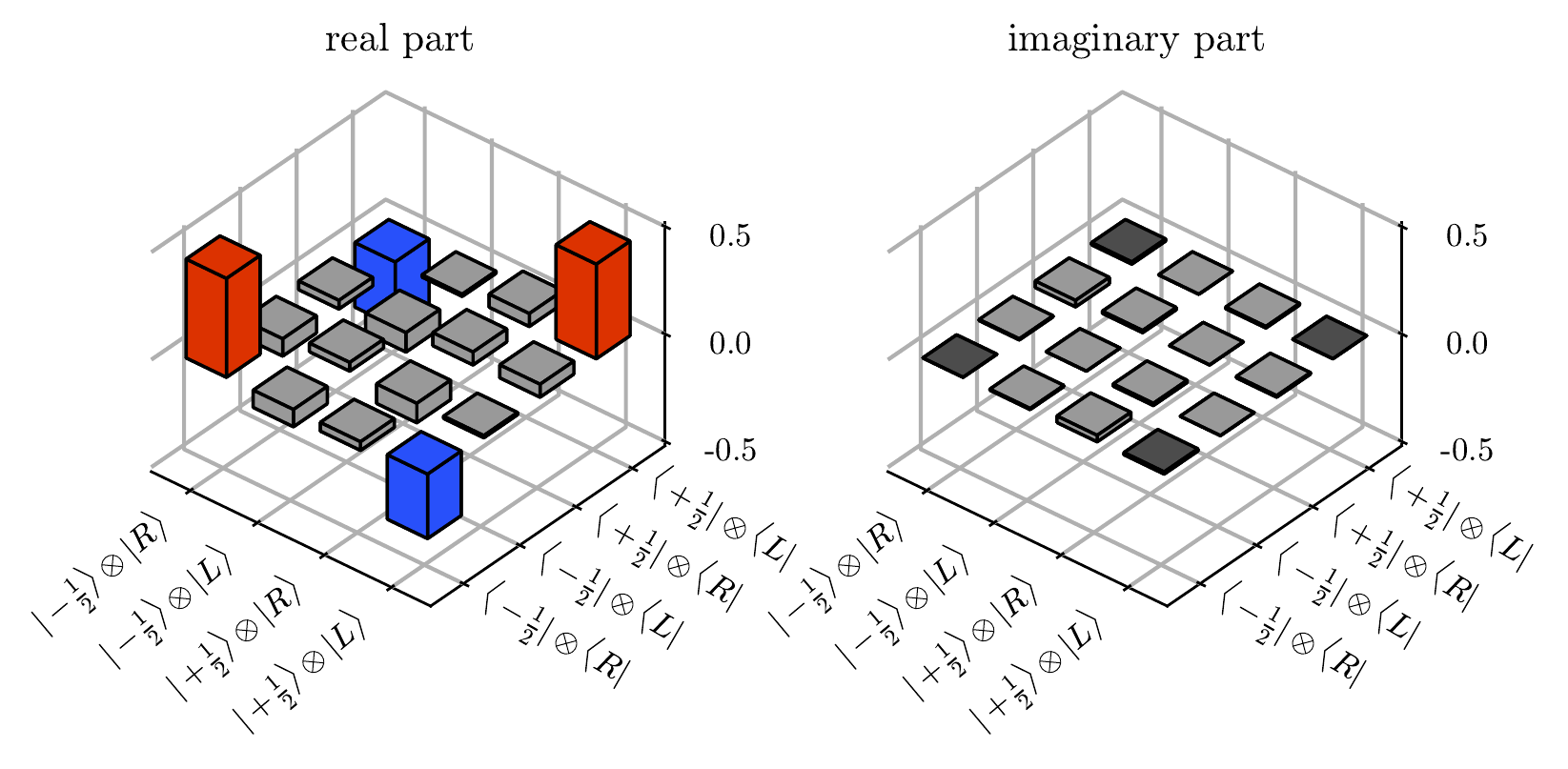}
	\caption{\label{fig.AtPhRho} Real and imaginary parts of the density matrices of the reconstructed atom-photon state for heralded absorption in the first (a) and second passage (b). Background and binning correction is applied.}
\end{figure}

\begin{table}
\caption{Fidelities and purities of the entanglement transfer. The values in brackets are corrected for background and binning.}
\label{tab:entanglementResults}
\begin{tabular}{c|c|c}
                                    & first passage & second passage \\
                                    \hline
    $\bra{\Psi^-}\rho\ket{\Psi^-}$ &  \SI{78.0(9)}{\percent} (\SI{82.4(10)}{\percent})  &   \\                                    
    $\bra{\Phi^-}\rho\ket{\Phi^-}$ &    &  \SI{52(1)}{\percent} (\SI{76(2)}{\percent}) \\    $\text{tr}(\rho^2)$ & \SI{65(1)}{\percent} (\SI{72(1)}{\percent})  & \SI{36(1)}{\percent} (\SI{64(5)}{\percent}) 
\end{tabular}
\end{table}

\section{Step\,2: Bell-state projection}
We now describe how a full Bell state measurement on an atom-photon state is facilitated by heralded absorption. We experimentally demonstrate and characterize the protocol by applying it for quantum state teleportation from an atom to a photon. 
Teleportation starts with a qubit encoded in the D$_{5/2}$ manifold, 
\begin{equation}
\ket{\varphi}_D = \alpha \ket{-\nicefrac{5}{2}}_D + \beta \ket{+\nicefrac{5}{2}}_D
\end{equation}
and the entangled two-photon state $\ket{\Psi^-}_{A,B}$. The Bell states between the \SI{854}{\nano\meter} photon in arm A 
and the atom in D$_{5/2}$ are defined by
\begin{equation}
\begin{split}
\ket{\Phi^\pm}_{A,D} = \frac{1}{\sqrt{2}} \left( \ket{R}_A \ket{-\nicefrac{5}{2}}_D \pm \ket{L}_A \ket{+\nicefrac{5}{2}}_D \right) \\
\ket{\Psi^\pm}_{A,D} = \frac{1}{\sqrt{2}} \left( \ket{R}_A \ket{+\nicefrac{5}{2}}_D \pm \ket{L}_A \ket{-\nicefrac{5}{2}}_D \right)
\end{split}
\label{AtPhBellStates}
\end{equation}
The joint state of all three qubits in the standard teleportation notation \cite{Bennett1993} is then 
\begin{equation}
\begin{split}
\ket{\Psi^-}_{A,B} \otimes \ket{\varphi}_D = \frac{1}{2} \Big( \ \ 
 &\left( \alpha \ket{R}_B - \beta \ket{L}_B \right) \otimes \ket{\Psi^+}_{A,D}\notag\\
-&\left( \alpha \ket{R}_B + \beta \ket{L}_B \right) \otimes \ket{\Psi^-}_{A,D}\notag\\
+&\left( \beta \ket{R}_B - \alpha \ket{L}_B \right) \otimes \ket{\Phi^+}_{A,D}\notag\\
-&\left( \beta \ket{R}_B + \alpha \ket{L}_B \right) \otimes \ket{\Phi^-}_{A,D}\ \Big)
\end{split}
\end{equation}
The Bell state projection on the atom and the photon in arm A is performed by the heralded absorption process, subsequent projection of the \SI{393}{\nano\meter} Raman photon (the herald) onto the linear polarizations of eq.\ (\ref{protocolLinPol}), and projection of the S$_{1/2}$ qubit state onto the superpositions of the Zeeman sub-levels
\begin{equation}
\begin{split}
\ket{+}_S =  \frac{\ket{-\nicefrac{1}{2}}_{S} + \ket{+\nicefrac{1}{2}}_{S}}{\sqrt{2}} \\
\ket{-}_S =  \frac{\ket{-\nicefrac{1}{2}}_{S} - \ket{+\nicefrac{1}{2}}_{S}}{i\,\sqrt{2}} 
\end{split}
\end{equation}
The operation as a Bell state measurement is understood by describing the eight possible measurement outcomes -- projection result of herald, projection result of atomic ground state, and absorption in first or second passage -- with the help of the Raman process operators for the first and second passage (eq.~\eqref{RamanOpFirst}, eq.~\eqref{RamanOpSecond}) 
\begin{equation}
\begin{split}
\left(\, \bra{H}_{393} \bra{+}_S\,\right) \, \hat{\text{R}}_{A,D}^{\text{\nth{1}}} &= \frac{1}{\sqrt{2}} \bra{\Phi^+}_{A,D} \\ 
\left(\, \bra{V}_{393} \bra{-}_S\,\right) \, \hat{\text{R}}_{A,D}^{\text{\nth{1}}} &= \frac{1}{\sqrt{2}} \bra{\Phi^+}_{A,D} \\
\end{split}
\label{BellProjectionPhiP}
\end{equation}
\begin{equation}
\begin{split}
\left(\, \bra{H}_{393} \bra{-}_S\,\right) \, \hat{\text{R}}_{A,D}^{\text{\nth{1}}} &= \frac{i}{\sqrt{2}} \bra{\Phi^-}_{A,D} \\
\left(\, \bra{V}_{393} \bra{+}_S\,\right) \, \hat{\text{R}}_{A,D}^{\text{\nth{1}}} &= \frac{-i}{\sqrt{2}} \bra{\Phi^-}_{A,D} \\
\end{split}
\label{BellProjectionPhiM}
\end{equation}
\begin{equation}
\begin{split}
\left(\, \bra{H}_{393} \bra{+}_S\,\right) \, \hat{\text{R}}_{A,D}^{\text{\nth{2}}} &= \frac{1}{\sqrt{2}} \bra{\Psi^+}_{A,D} \\ 
\left(\, \bra{V}_{393} \bra{-}_S\,\right) \, \hat{\text{R}}_{A,D}^{\text{\nth{2}}} &= \frac{1}{\sqrt{2}} \bra{\Psi^+}_{A,D} \\
\end{split}
\label{BellProjectionPsiP}
\end{equation}
\begin{equation}
\begin{split}
\left(\, \bra{H}_{393} \bra{-}_S\,\right) \, \hat{\text{R}}_{A,D}^{\text{\nth{2}}} &= \frac{i}{\sqrt{2}} \bra{\Psi^-}_{A,D} \\
\left(\, \bra{V}_{393} \bra{+}_S\,\right) \, \hat{\text{R}}_{A,D}^{\text{\nth{2}}} &= \frac{-i}{\sqrt{2}} \bra{\Psi^-}_{A,D} \\
\end{split}
\label{BellProjectionPsiM}
\end{equation}
One sees that we project onto the $\ket{\Phi^\pm}_{A,D}$ states in the first passage and onto the $\ket{\Psi^\pm}_{A,D}$ states in the second passage, hence all four Bell states are distinguished. The result of the projection measurement is the classical 2-bit information needed to complete the quantum state teleportation onto photon B. 

\begin{figure}[htb]
	\hfill
	\begin{minipage}[t]{4.25cm}
		\includegraphics[width=4.25cm]{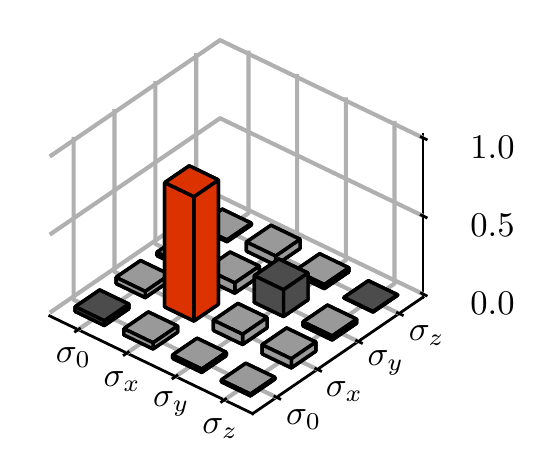}

		(a)\hspace{5mm} $\ket{\Phi^-}$
		
		\includegraphics[width=4.25cm]{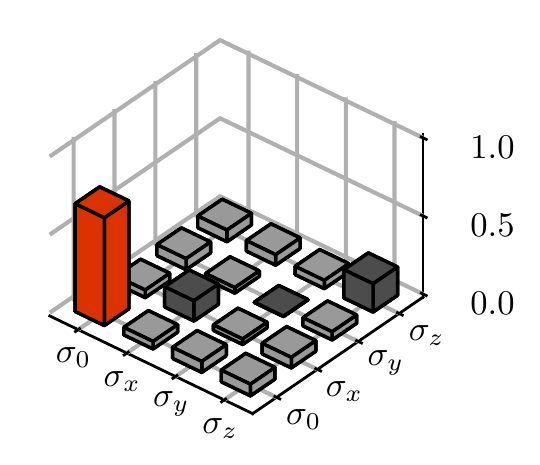}
        
		(c)\hspace{5mm} $\ket{\Psi^-}$		
	\end{minipage}
	\hfill
	\begin{minipage}[t]{4.25cm}
		\includegraphics[width=4.25cm]{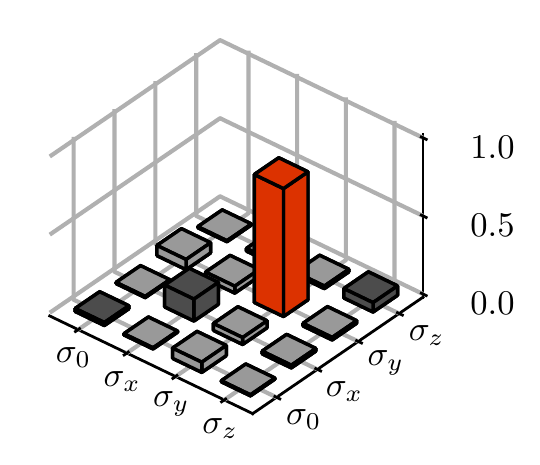}
	
		(b)\hspace{5mm} $\ket{\Phi^+}$

		\includegraphics[width=4.25cm]{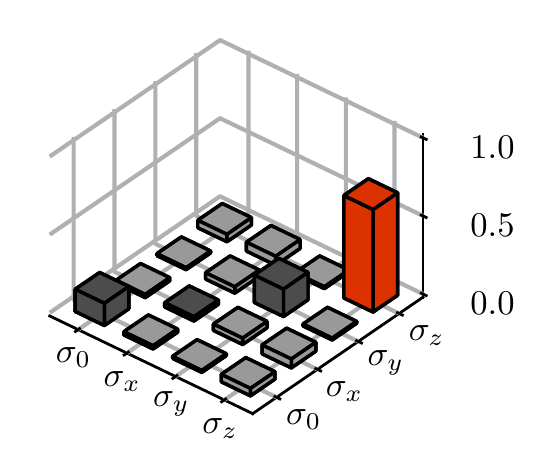}
    
		(d)\hspace{5mm} $\ket{\Psi^+}$
	\end{minipage}
	\hfill\mbox{}
	\caption{\label{fig.teleport_PM_abschi} Reconstructed process matrices of the quantum state teleportation conditioned on the Bell-state measurement results. (a) and (b) correspond to the $\ket{\Phi^-}$ and $\ket{\Phi^+}$ Bell states, (c) and (d) correspond to the $\ket{\Psi^-}$ and $\ket{\Psi^+}$ Bell states. Background and binning correction is applied.}
\end{figure}

To demonstrate the full Bell-state measurement, we perform teleportation of various input qubit states: we prepare the atomic basis states $\ket{-\nicefrac{5}{2}}_D$ and $\ket{+\nicefrac{5}{2}}_D$, as well as various superpositions $\left( \ket{-\nicefrac{5}{2}}_D + e^{i\, \phi} \ket{+\nicefrac{5}{2}}_D \right)/\sqrt{2}$. For variation of the phase $\phi$, we take advantage of the Larmor precession in the atom; the phase is well defined in every absorption event through the time between preparation of the atom and detection of the \SI{393}{\nano\meter} herald \cite{Kurz2016}. Teleportation events are binned according to $\phi$ for evaluation. 
Using all successful runs of the teleportation protocol (detection of herald and photon B), state tomography on the \SI{854}{\nano\meter} target photon B is applied. It consists in measuring the photon in the H/V (horizontal/vertical), D/A (diagonal/anti-diagonal), and R/L (right/left circular) polarization bases, successively, for each atomic input state. The measurements are then used to reconstruct the quantum process matrix \cite{Chuang1997} in the Pauli basis ${\sigma_0, \sigma_x, \sigma_y, \sigma_z}$ with a maximum likelihood algorithm. The result of the reconstruction is shown in figure \ref{fig.teleport_PM_abschi}. Conditioned on the eight possible outcomes of eqs.~\eqref{BellProjectionPhiP}-\eqref{BellProjectionPsiM}, we see that either a $\sigma_x$ (figure \ref{fig.teleport_PM_abschi}(a)), a $\sigma_y$ (figure \ref{fig.teleport_PM_abschi}(b)), 
no rotation (figure \ref{fig.teleport_PM_abschi}(c)) or $\sigma_z$ (figure \ref{fig.teleport_PM_abschi}(d))  must be applied to the target photon to reveal the prepared input state. The corresponding entries in the process matrix are referred to as process fidelities and listed in table~\ref{tab:processfidelities}. The mean process fidelity is \SI{76 +- 9 }{\percent} (\SI{81 +- 5}{\percent} with background and binning correction). 
Like before, the most relevant contribution to the background is caused by accidental coincidences of the SPDC source. The Bell state projection is therefore benchmarked by the background-corrected value, but we would have to take the uncorrected value when quantifying the teleportation fidelity with the used experimental apparatus. 

\begin{table}
\caption{Process fidelities of atom-to-photon teleportation for the four Bell measurement results, with and without correction. }
\label{tab:processfidelities}
\begin{tabular}{c | c | c}
state & fidelity w correction & fidelity w/o correction \\
\hline
$\ket{\Phi^-}$ & \SI{84 +- 8}{\percent} & \SI{83(8)}{\percent}\\
$\ket{\Phi^+}$ & \SI{86(7)}{\percent} & \SI{85(7)}{\percent}\\
$\ket{\Psi^-}$ & \SI{77(12)}{\percent} & \SI{66(6)}{\percent}\\
$\ket{\Psi^+}$ & \SI{76(8)}{\percent} & \SI{69(5)}{\percent}
\end{tabular}
\end{table}

\section{Discussion and Conclusion}\label{sec.conclusion}

The statistics of Fig.~\ref{fig.teleport_PM_abschi} and Table~\ref{tab:processfidelities} are based on \num{511670886} executions of the teleportation protocol (i.e., coincident registration of the absorption herald and the partner photon) in 49.75~h hours of exposure. 

Given the free-space coupling between ion and in- and outgoing photons, the absorption probability for an incoming photon is  \num{1.04e-3} in the first passage, and \num{1.32e-04} in the second passage. The measured detection efficiency for the absorption herald at 393\,nm is 1.64\,\%. This leads to a success probability for an individual heralded absorption event of \num{1.71e-5} in the first passage, and \num{2.16e-06} in the second passage (see the Supplement for more details). Much higher rates may be obtained when absorption and release of the herald are enhanced by optical resonators \cite{Schupp2021,Brekenfeld2020}. Nevertheless, the fidelity of the protocol is independent of the success probability, because the herald allows us to filter out successful events. Factors that limit the fidelity are polarization impurities, atomic decoherence due to magnetic noise, inherent background from the SPDC process, and the binning of the detected photon arrival times. The data of Fig.~\ref{fig.teleport_PM_abschi} and the fidelity numbers in the middle column of Table~\ref{tab:processfidelities}
are corrected for background and binning. We characterized the impact of magnetic field fluctuations by evaluating data only from the first \SI{50}{\micro\second} of the total \SI{350}{\micro\second} exposure window in the state mapping procedure. The fidelity for the first passage, $\bra{\Psi^-}\rho\ket{\Psi^-}$ (see table~\ref{tab:entanglementResults}), increases from \SI{78.0(9)}{\percent} to \SI{84(2)}{\percent} (\SI{82.4(10)}{\percent} to \SI{89(2)}{\percent}, with corrections). For the second passage, however, using \SI{50}{\micro\second} exposure time leads to too large statistical uncertainties, such that we used \SI{350}{\micro\second} throughout. 
The remaining infidelity, with respect to what one expects from the initial two-photon state (Fig.~\ref{fig.PhRho}), originates predominantly from accumulated polarization impurities in the optical paths and phase drifts of the SPDC source on long time scales.

In conclusion, we presented a proof-of-concept experiment that points a way to a full four-state Bell measurement on two incoming photons, by employing a single quantum memory. We demonstrated and characterized two parts of the overall protocol separately. In the first step, a single photon is mapped onto a single-atom memory qubit; the procedure is verified by the faithful transfer of two-photon entanglement to memory-photon entanglement. In the the second step, the full Bell-state measurement on the 2-qubit state between memory and another photon is demonstrated. It is implemented by using heralded absorption and discriminating the four output states between herald and final atomic qubit. We verified this second step by applying it in quantum state teleportation from the memory qubit onto the partner photon of the absorbed one, using the SPDC photon pair source as resource of entanglement. The presented experiments confirm that, by the use of heralded operations, high-fidelity quantum operations may be realized despite possibly low success probabilities \cite{Bock2018, Lenhard2015, Kurz2014}. 

While the overall protocol offers a means to overcome the fundamental efficiency limitation of 50\,\% of a 2-photon Bell state measurement with linear optics, the real efficiency of the presented implementation is severely limited by other factors. As already mentioned, this situation changes when using optical resonators. The approach of \cite{Brekenfeld2020}, using separate resonators for the incoming- and heralding photon, reaches already a single-photon heralding efficiency of 11(1)\,\%, and is importantly not limited by fundamental factors as well as the described protocol. We note two further features of our protocol. Firstly, the two photons need not interfere directly with each other; as a consequence they may be distinguishable in terms of temporal or spectral properties and need not be Fourier-limited. Secondly, the approach works with consecutively arriving photons, \textit{i.e.}, the required arrival time distribution is relaxed by the protocol; importantly, it may be used as an asynchronous Bell measurement in a quantum repeater segment \cite{Loock2020}.

\section{Methods}

\subsection{Atomic manipulations}
A set of \SI{397}{\nano\meter} and \SI{866}{\nano\meter} lasers is used for Doppler cooling on the S$_{1/2}$-P$_{1/2}$ transition, and for fluorescence detection. Coherent manipulations on the D$_{5/2}$ and S$_{1/2}$ Zeeman sub-levels of the ion are performed with a radio-frequency (RF) magnetic field antenna and a narrow-band \SI{729}{\nano\meter} laser. 

The ion is first initialized in the pure state $\ket{-\nicefrac{1}{2}}_S$ by Doppler cooling and frequency-selective optical pumping. A resonant $\pi/2$ RF pulse creates a coherent superposition of the S$_{1/2}$ Zeeman sub-levels. To generate coherent superposition of the D$_{5/2}$ Zeeman states, two \SI{729}{\nano\meter} $\pi$-pulses transfer the populations coherently  from $\ket{-\nicefrac{1}{2}}_S$ to $\ket{-\nicefrac{5}{2}}_D$ and from $\ket{+\nicefrac{1}{2}}_S$ to $\ket{+\nicefrac{5}{2}}_D$. Additional details are provided in \cite{Kurz2014, Kurz2016}.

\subsection{Ion-photon quantum state reconstruction}
\label{sec:methods-state-reconstruction}
We perform a tomographic complete set of measurements to reconstruct the two-qubit ion-photon quantum state after heralded absorption, $\rho$. The chosen observables are the 16 tensor products 
$\lbrace \hat{\sigma}_i \otimes \hat{\sigma}_j \;|\; i,j = 0,\ldots,3 \rbrace$ of identity and Pauli spin operators, $\lbrace \hat{\sigma}_{0,\ldots,3}\rbrace = \lbrace \hat{1} , \hat{\sigma}_x , \hat{\sigma}_y , \hat{\sigma}_z \rbrace$. Their expectation values 
\begin{equation}
\langle \hat{\sigma}_i \otimes \hat{\sigma}_j \rangle_\rho = \textrm{tr} \left( \hat{\sigma}_i \otimes \hat{\sigma}_j \ \rho \right) 
\end{equation}
allow for direct linear reconstruction of $\rho$ according to 
\begin{equation}
\rho = \frac{1}{4} \sum_{i,j=0}^{3} \left\langle \hat{\sigma}_i \otimes \hat{\sigma}_j \right\rangle_\rho \cdot \  \hat{\sigma}_i \otimes \hat{\sigma}_j
\end{equation}  
We write the expectation values as
\begin{equation}
\begin{split}
\left\langle  \hat{\sigma}_i \otimes \hat{\sigma}_j \right\rangle 
&= \lambda_{\ket{0_i}}  \cdot  P\left( \ket{0_i} \right)  \cdot \left\langle  \hat{\sigma}_j \right\rangle\Big|_{\ket{0_i}} \\
&+ \lambda_{\ket{1_i}}  \cdot  P\left( \ket{1_i} \right)  \cdot \left\langle  \hat{\sigma}_j \right\rangle\Big|_{\ket{1_i}}
\end{split}
\end{equation}
where $\lambda_{\ket{0_i}}$, $\lambda_{\ket{1_i}}$ denote the eigenvalues of the observables $\hat{\sigma}_i$ of the photonic qubit, $P\left( \ket{0_i} \right)$ and $P\left( \ket{1_i} \right)$ are the probabilities to detect the photonic qubit in the eigenstates $\ket{0_i}$ and $\ket{1_i}$ of this observable, and
$\left\langle\hat{\sigma}_j \right\rangle\Big|_{\ket{0_i}}$ and $\left\langle\hat{\sigma}_j \right\rangle\Big|_{\ket{1_i}}$ are the conditioned expectation values of the atomic qubit after detection of the photonic qubit in the eigenstate $\ket{0_i}$ respectively $\ket{1_i}$.

Photonic projection in the basis settings $\ket{H}$/$\ket{V}$,  $\ket{D}$/$\ket{A}$, and $\ket{R}$/$\ket{L}$ is performed with a set of quarter-wave and half-wave plates and two APDs at the outputs of a polarizing beam splitter. For atomic projection onto the $\ket{\pm\nicefrac{1}{2}}_{S}$ basis, i.e. for $\left\langle\hat{\sigma}_z\right\rangle$, we perform electron shelving and fluorescence detection. An additional $\nicefrac{\pi}{2}$ RF pulse before electron shelving is used to project onto the superposition basis. Atomic superpositions carry a Larmor precession phase defined by the time instant at which the heralded absorption happened. We therefore plot the projection result after the $\nicefrac{\pi}{2}$ pulse as a function of this Larmor phase and extract $\langle\hat{\sigma}_x\rangle$ and $\langle\hat{\sigma}_y\rangle$ from a fit to a sinusoidal function 
\begin{equation}
    \frac{V}{2}\sin(x-\phi_0) + \frac{1}{2}~~.
    \label{Larmor-fit-fctn}
\end{equation}

\subsection{Binning correction}
\label{sec:methods-binning-correction}

In order to correct for the influence of the binning, we change the fit function from Eq.~(\ref{Larmor-fit-fctn}) to
\begin{equation}
     \frac{V}{2} \frac{N}{\pi} \sin\left(\frac{\pi}{N}\right) \sin(x - \phi_0) + \frac{1}{2}
\end{equation}
with the number of bins $N$.

\subsection{Larmor precession and spin echo}

Due to the Zeeman splitting in the static magnetic field, the atomic superposition states undergo Larmor precession. At our magnetic field strength, the D$_{5/2}$ superposition precesses with \SI{24}{\mega\hertz}, while the S$_{1/2}$ qubit oscillates at \SI{8}{\mega\hertz}. We take advantage of the different Larmor precession frequencies and the variable waiting time until heralded absorption occurs, to provide atomic superposition states with different phases in each run. Our phase reference is an RF oscillator set to the Larmor frequency of the ground-state qubit. The phase of each event is calculated a posteriori from the arrival time of the absorption herald. 

Fluctuations in the ambient magnetic field change the Larmor frequencies and give rise to phase errors. While slow variations (\SI{50}{Hz} and harmonics) are suppressed using a feed-forward stabilization via compensation coils, faster noise leads to decoherence of the atomic qubits. A spin-echo \cite{Hahn1950} technique compensates these errors partially: by applying a $\pi$-pulse on the ground-state qubit and a subsequent waiting time $\tau_S = 3\times\tau_D$ before projecting the atomic state, the phase error accumulated in D$_{5/2}$ during the time $\tau_D$ between preparation and absorption is corrected. The factor 3 in the waiting time is determined by the ratio between the Larmor frequencies of the D$_{5/2}$ and the S$_{1/2}$ Zeeman qubits.

Instead of triggering the spin-echo pulses directly by the herald, we check every 500~ns whether absorption has occured. By choosing the loop time, and thereby also $\tau_D$ and $\tau_S$, to be an integer multiple of the difference of the two Larmor frequencies, we avoid fractional phase accumulation. 

\section{Acknowledgements}
\begin{acknowledgments}
We acknowledge support by the German Federal Ministry of Education and Research (BMBF) through projects Q.Link.X (16KIS0864), CaLas (13N14908), and QR.X (16KISQ001K). 
\end{acknowledgments}

\bibliography{bibliothek}

\end{document}

\title{Supplementary information: Full Bell-basis measurement of an atom-photon 2-qubit state and its application for quantum networks}

\author{Elena Arensk\"otter}
\thanks{} 
\affiliation{Experimentalphysik, Universit\"at des Saarlandes, 66123 Saarbr\"ucken, Germany}

\author{Stephan Kucera}
\thanks{}
\affiliation{Experimentalphysik, Universit\"at des Saarlandes, 66123 Saarbr\"ucken, Germany}

\author{Omar Elshehy}
\thanks{}
\affiliation{Experimentalphysik, Universit\"at des Saarlandes, 66123 Saarbr\"ucken, Germany}

\author{Max Bergerhoff}
\thanks{}
\affiliation{Experimentalphysik, Universit\"at des Saarlandes, 66123 Saarbr\"ucken, Germany}

\author{Matthias Kreis}
\thanks{}
\affiliation{Experimentalphysik, Universit\"at des Saarlandes, 66123 Saarbr\"ucken, Germany}

\author{Léandre Brunel}
\thanks{Current address: Department of Physics, University of Virginia, Charlottesville, Virginia 22903, USA}
\affiliation{Experimentalphysik, Universit\"at des Saarlandes, 66123 Saarbr\"ucken, Germany}

\author{J\"urgen Eschner}
\thanks{\href{mailto:juergen.eschner@physik.uni-saarland.de}{juergen.eschner@physik.uni-saarland.de}}
\affiliation{Experimentalphysik, Universit\"at des Saarlandes, 66123 Saarbr\"ucken, Germany}

\date{\today}% It is always \today, today,

\maketitle

\section{Single-photon to atom quantum state mapping}

By pumping the SPDC resonator from only one side, we generate orthogonally polarized photon pairs, that are split on the polarizing beam splitter. We use the photon in arm A as the herald for a single photon in arm B, that is sent to the ion. The polarization of photon B is rotated to one of the 6 input polarizations $\ket{i}$~$\in$~\{$\ket{H}$, $\ket{D}$, $\ket{V}$, $\ket{A}$, $\ket{R}$, $\ket{L}$\}. We reconstruct the process matrix \cite{Chuang1997} by quantum state tomography on the final atomic state $\rho_{f,i}$ for each input polarization. We correct the background for lost-partner events and detector dark counts.

\begin{figure}[htb]
	\centering
	\includegraphics [width=0.45\mycolumnwidth]{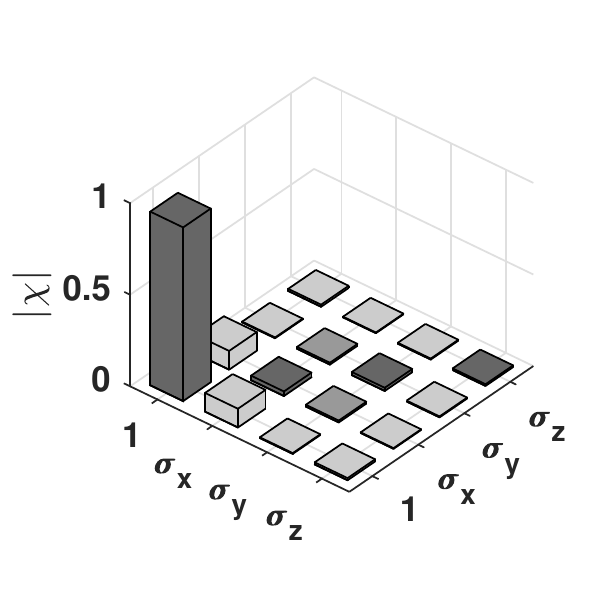}
	\caption{\label{fig.HerAbsChi} Absolute values of the reconstructed process matrix with the identity part of \SI{96.2}{\percent} and mean overlap fidelity of \SI{96.7(8)}{\percent}.}
	% Alte Messung von Stephan, 4-Spiegel Quelle
\end{figure}

Figure \ref{fig.HerAbsChi} shows the absolute values of the process matrix $\chi$ with the identity part of $\chi_{11} = \SI{96.2}{\percent}$ identified as the process fidelity. From this value, we infer the mean overlap fidelity $\mean{F} = (2\,\chi_{11} + 1)/3 = \SI{96.6}{\percent}$ (\SI{93.3}{\percent} without background correction), which agrees with the mean of measured overlap fidelities of $1/6 \, \sum_i \bra{i}\rho_{f,i}\ket{i} = \SI{96.7(8)}{\percent}$ (\SI{93(3)}{\percent} without background correction). The major limitation is the coherence time of the ion due to magnetic field fluctuations during the exposure time (\SI{56}{\micro\second} in Fig.~\ref{fig.HerAbsChi}).
We repeated the protocol \num{1.13e8} times for the measurement, which led to a total exposure time of \SI{1}{\hour}\SI{45}{\minute}. In this time, we generated \num{9.3e8} heralded fibre-coupled \SI{854}{\nano\meter} photons, and we recorded \num{7810} detected (background corrected) coincidences between the \SI{854}{\nano\meter} herald and the emitted \SI{393}{\nano\meter} photon. We extract the state mapping probability for a fiber coupled photon to be $\eta = \num{8.4e-6}$. Efficiencies for the entanglement transfer and the teleportation differ from these values and are discussed in the following sections. %For the following experiments, we increased the detection efficiency of the \SI{393}{\nano\meter}; this leads to a success probability of $\eta = \num{2.1e-5}$.

\section{Polarisation control}
For setting the polarisation of photon B, we adjust the measurement basis at the detectors: we send laser light backwards through our setup and detect it with a polarimeter via a flip mirror. The rotation matrix of this flip mirror is compensated by three waveplates (quarter-wave, half-wave, quarter-wave). The projection setup generates 37 different input polarizations, equally distributed over the Poincare sphere. From the measured polarizations we then calculate the rotation matrix $M_B$.

The polarisation between the source and the trap is, in contrast, compensated behind the source: we insert a 99:1 fiber beam splitter in the fiber connecting source and trap. Between the source and the beam splitter we install a fiber polarisation controller (PolaRITE III PCD-M02-854), and between beam splitter and trap a 3-paddle polarisation controller. The 1 \% output of the beam splitter is connected to a reference polarimeter. For calibration, we send a reference laser beam from the back-reflection direction. Its polarization is first adjusted to $\ket{R}$ at the position of the ion, indicated by maximal suppression of the corresponding orthogonal transition. With an optional quarter-wave plate the reference polarization is rotated to $\ket{H}$. The next step is to compensate the polarisation to the reference polarimeter with the fiber paddles. In the next step we compensate the polarisation to the source, which is measured with an additional polarimeter in front of the source. The last step is to send the light from the source direction and compensate the polarisation of the back-reflection, measured again with the reference polarimeter.

\section{Photon-photon to atom-photon entanglement transfer: data and evaluation}

\begin{figure}[htb]
	\centering
	\begin{subfigure}{0.45\mycolumnwidth}
		\includegraphics[width=\textwidth]{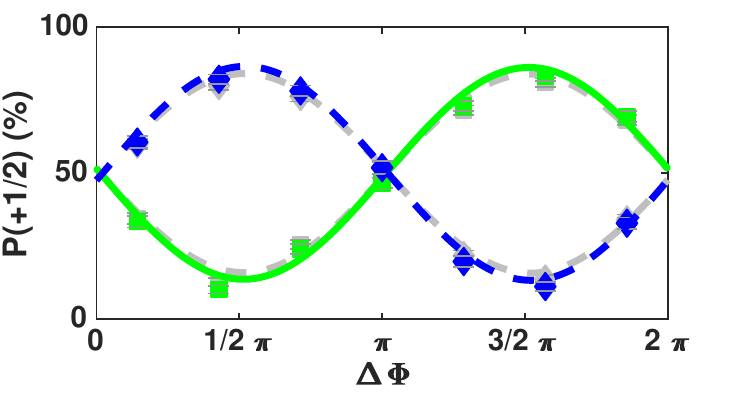}
		\caption{}
	\end{subfigure}
	\begin{subfigure}{0.45\mycolumnwidth}	
		\includegraphics[width=\textwidth]{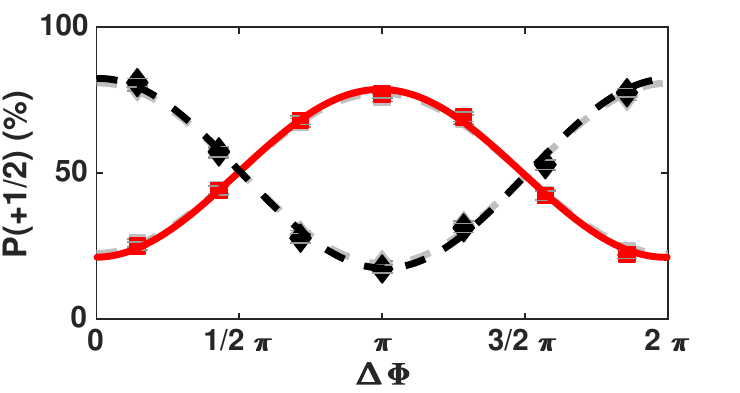}
		\caption{}
	\end{subfigure}
	
	\begin{subfigure}{0.45\mycolumnwidth}
		\includegraphics[width=\textwidth]{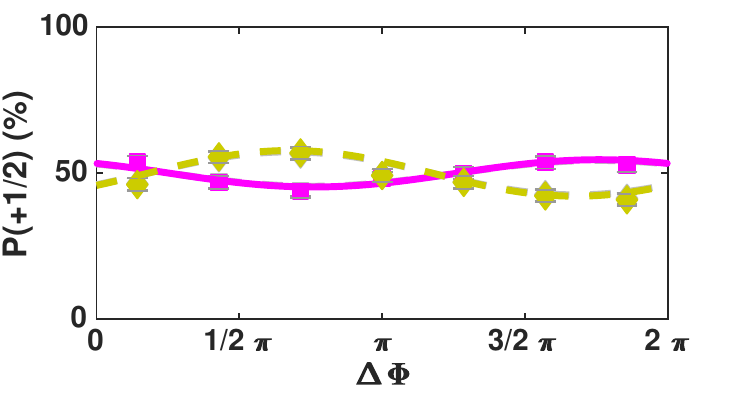}
		\caption{}
	\end{subfigure}
	\begin{subfigure}{0.45\mycolumnwidth}	
		\includegraphics[width=\textwidth]{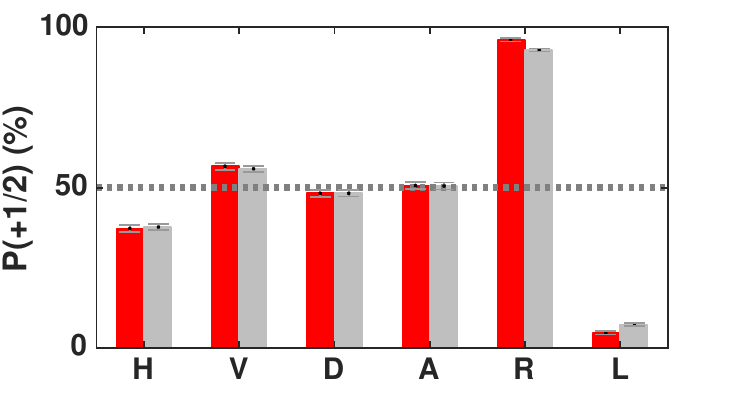}
		\caption{}
	\end{subfigure}
	
	\caption{\label{fig.EntTransDataPlotsFirst}
		(a-c) Fringes (with basis rotation) for the six projector settings of the first passage and the \SI{854}{\nano\meter} partner photon. The gray dashed lines show the data without correction:
		(a)~green~(squares):~$\ket{H}$,~blue~(diamonds):~$\ket{V}$
		(b)~red~(squares):~$\ket{D}$,~black~(diamonds):~$\ket{A}$
		(c)~magenta~(squares):~$\ket{R}$,~yellow~(diamonds):~$\ket{L}$
		(d)~Probabilities (without $\nicefrac{\pi}{2}$ RF pulse) for the six projector settings of the \SI{854}{\nano\meter} partner photon (red bars). The gray bars are without correction.
	}
\end{figure}

\begin{figure}[htb]
	\centering
	\begin{subfigure}{0.45\mycolumnwidth}
		\includegraphics[width=\textwidth]{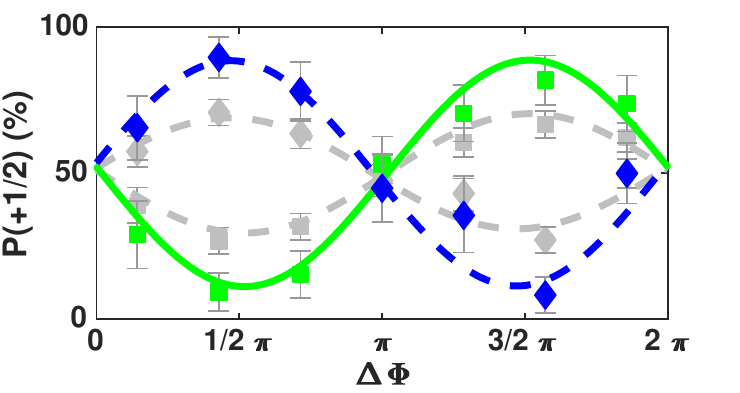}
		\caption{}
	\end{subfigure}
	\begin{subfigure}{0.45\mycolumnwidth}	
		\includegraphics[width=\textwidth]{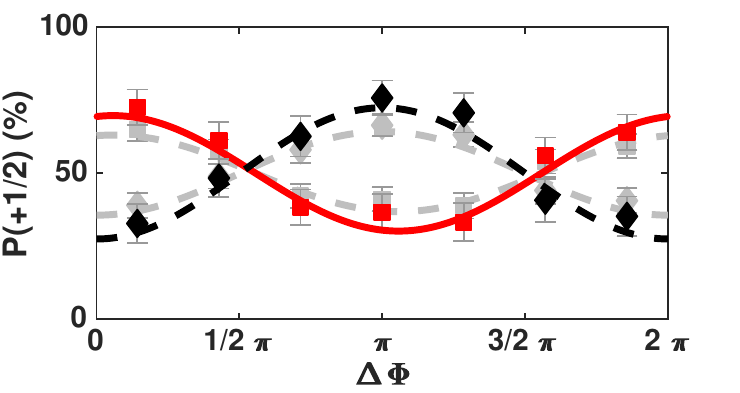}
		\caption{}
	\end{subfigure}
	
	\begin{subfigure}{0.45\mycolumnwidth}
		\includegraphics[width=\textwidth]{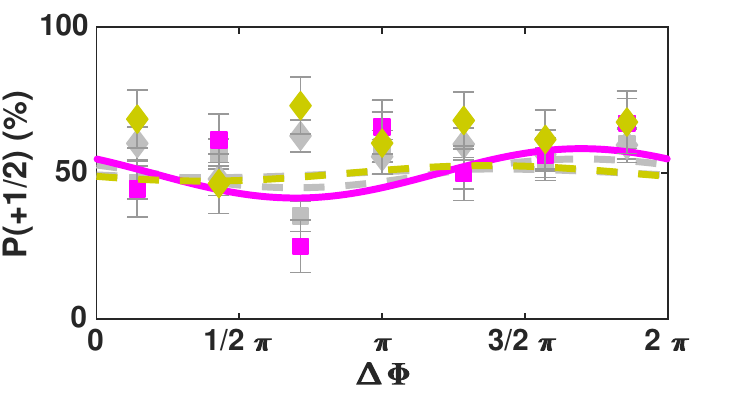}
		\caption{}
	\end{subfigure}
	\begin{subfigure}{0.45\mycolumnwidth}	
		\includegraphics[width=\textwidth]{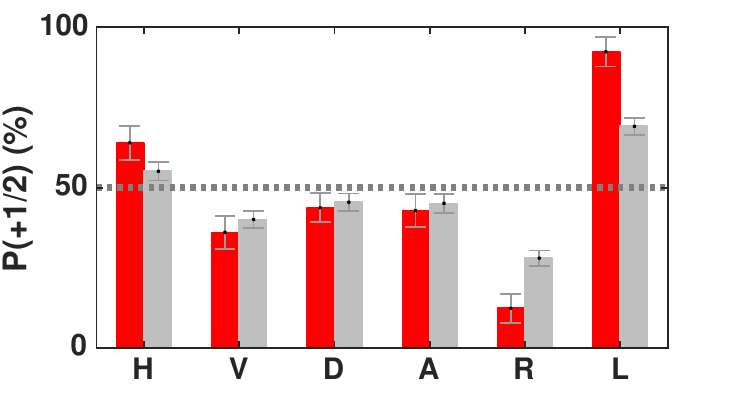}
		\caption{}
	\end{subfigure}
	
	\caption{\label{fig.EntTransDataPlotsSecond}
		(a-c) Fringes (with basis rotation) for the six projector settings of the second passage and the \SI{854}{\nano\meter} partner photon. The gray dashed lines show the data without correction:
		(a)~green~(squares):~$\ket{H}$,~blue~(diamonds):~$\ket{V}$
		(b)~red~(squares):~$\ket{D}$,~black~(diamonds):~$\ket{A}$
		(c)~magenta~(squares):~$\ket{R}$,~yellow~(diamonds):~$\ket{L}$
		(d)~Probabilities (without basis rotation) for the six projector settings of the \SI{854}{\nano\meter} partner photon (red bars). The gray bars are without correction.
	}
\end{figure}

Figures \ref{fig.EntTransDataPlotsFirst} and \ref{fig.EntTransDataPlotsSecond} show the atomic state analysis under condition of the six different \SI{854}{\nano\meter} projector settings with and without $\nicefrac{\pi}{2}$ RF pulse. The gray fringes and bars are without background correction. Out of this data set we calculate the conditioned expectation values of the atomic qubit. 

\FloatBarrier

\section{Quantum state teleportation: data and evaluation}
\begin{figure*}[htb]
		\centering
		\begin{tabular}{lc||c|c|c|c|}
			& & $\ket{\Phi^-}$ & $\ket{\Phi^+}$ & $\ket{\Psi^-}$ & $\ket{\Psi^+}$  \\ 
			\hline 
			\hline
			\multirow{3}*{\rotatebox{90}{$\left( \ket{-\nicefrac{5}{2}}_D + e^{i\, \phi} \ket{+\nicefrac{5}{2}}_D \right)/\sqrt{2}$}} &  
			\rotatebox{90}{$\qquad\quad\ket{H}_{\rph}$} &
			  \fboxrule0pt\fboxsep4pt\fbox{\includegraphics[width=0.2\textwidth]{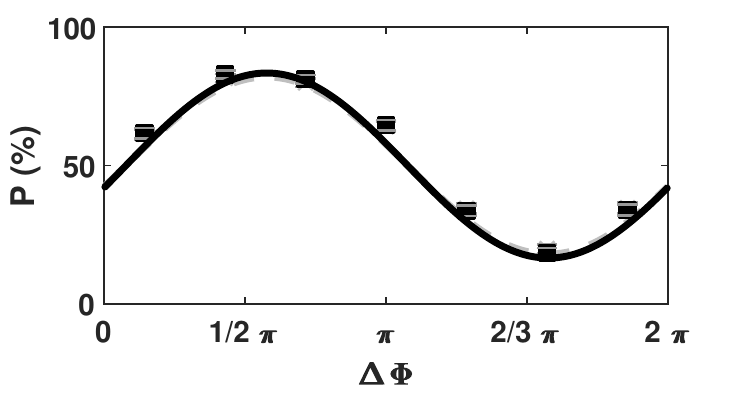}} &
			  \fboxrule0pt\fboxsep4pt\fbox{\includegraphics[width=0.2\textwidth]{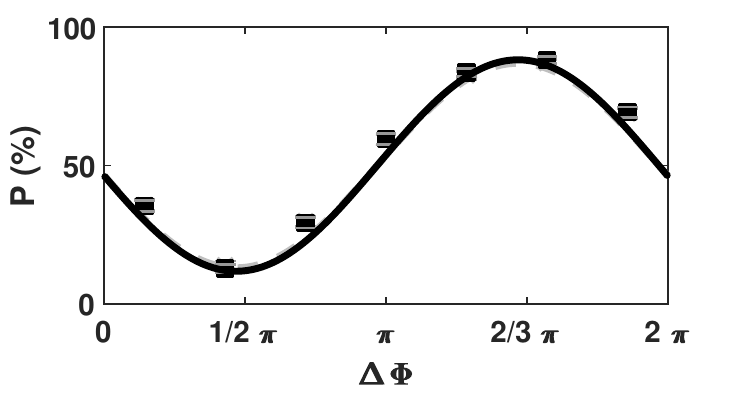}} & 
			  \fboxrule0pt\fboxsep4pt\fbox{\includegraphics[width=0.2\textwidth]{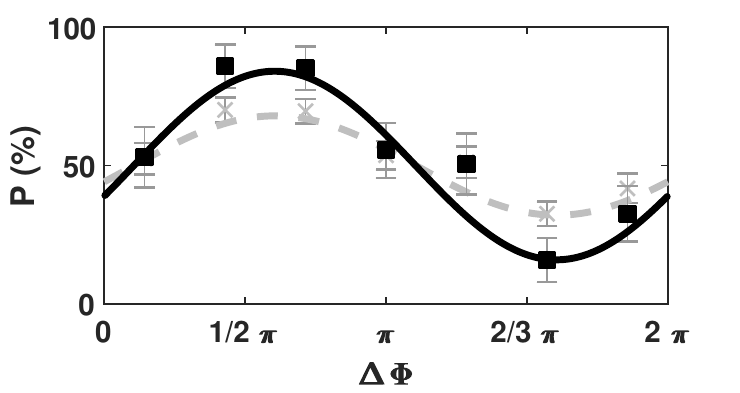}} & 
			  \fboxrule0pt\fboxsep4pt\fbox{\includegraphics[width=0.2\textwidth]{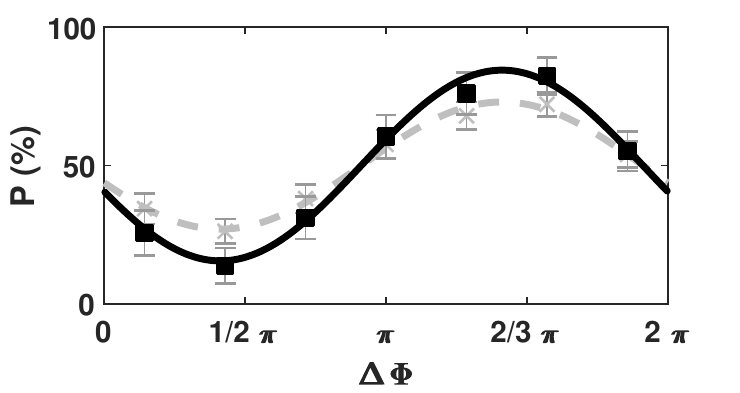}} \\ 
		    & \rotatebox{90}{$\qquad\quad\ket{D}_{\rph}$} &  \fboxrule0pt\fboxsep4pt\fbox{\includegraphics[width=0.2\textwidth]{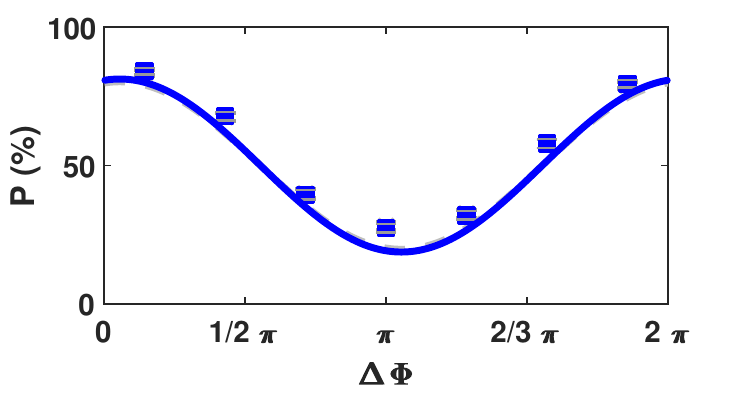}} &
			  \fboxrule0pt\fboxsep4pt\fbox{\includegraphics[width=0.2\textwidth]{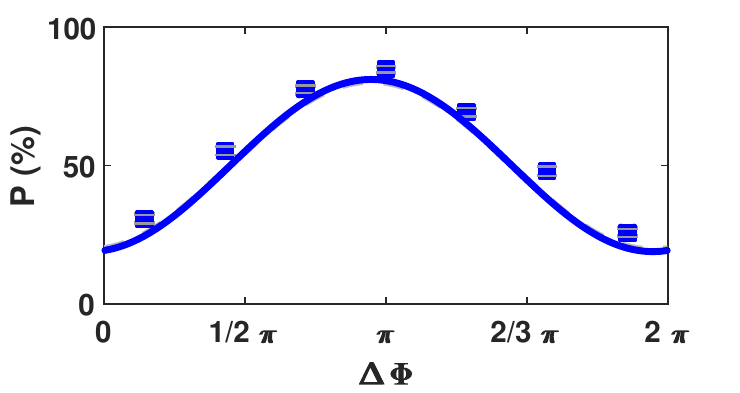}} & 
			  \fboxrule0pt\fboxsep4pt\fbox{\includegraphics[width=0.2\textwidth]{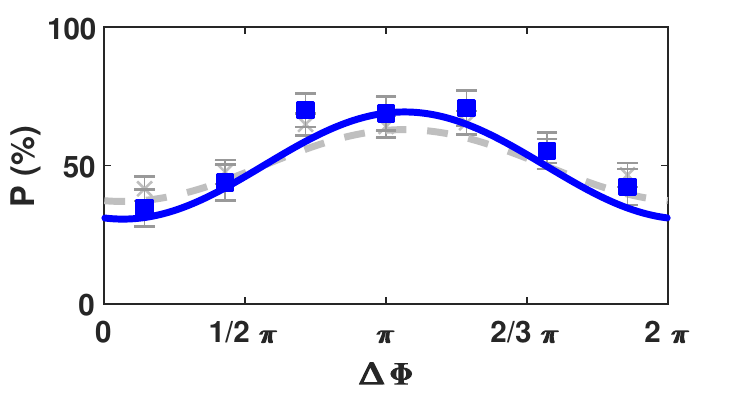}} & 
			  \fboxrule0pt\fboxsep4pt\fbox{\includegraphics[width=0.2\textwidth]{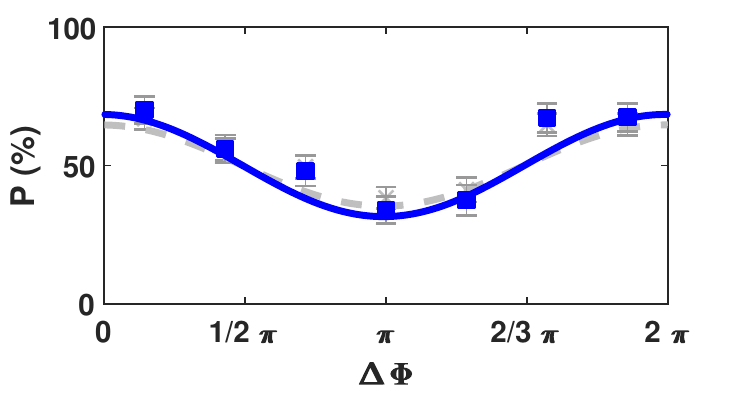}} \\ 
			& \rotatebox{90}{$\qquad\quad\ket{R}_{\rph}$}&  \fboxrule0pt\fboxsep4pt\fbox{\includegraphics[width=0.2\textwidth]{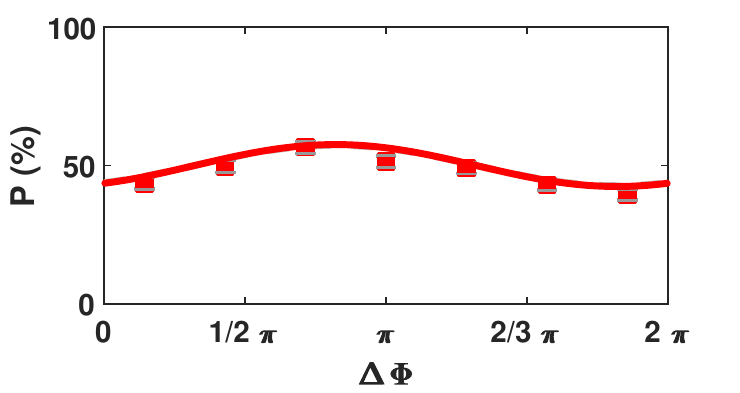}} &
			  \fboxrule0pt\fboxsep4pt\fbox{\includegraphics[width=0.2\textwidth]{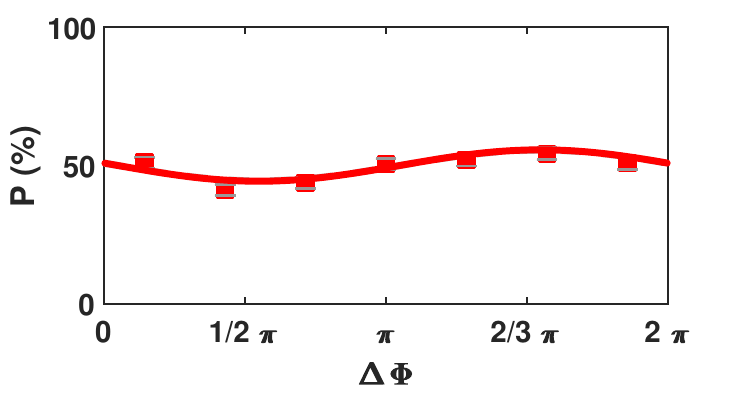}} & 
			  \fboxrule0pt\fboxsep4pt\fbox{\includegraphics[width=0.2\textwidth]{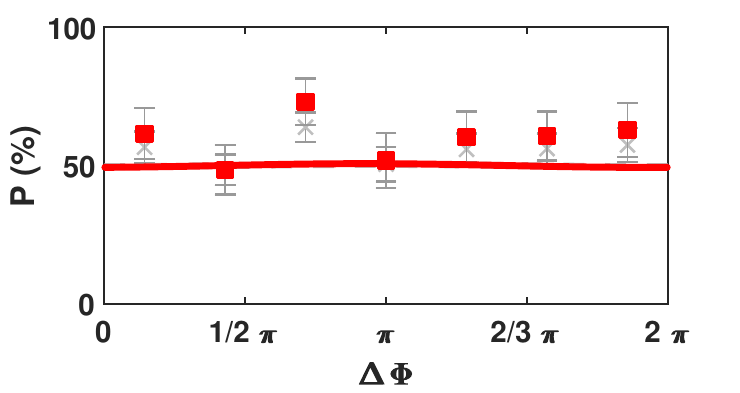}} & 
			  \fboxrule0pt\fboxsep4pt\fbox{\includegraphics[width=0.2\textwidth]{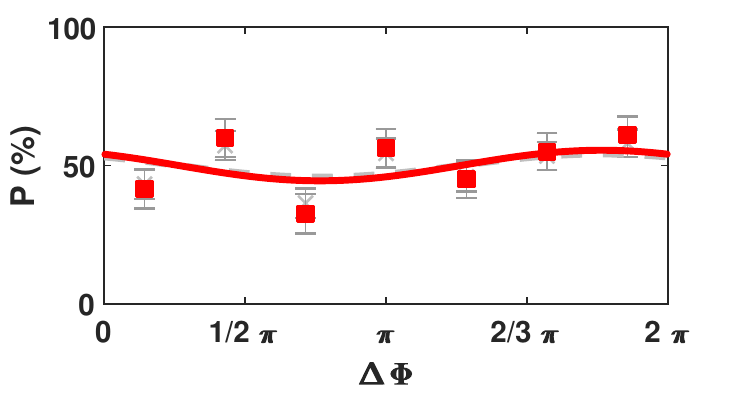}} \\ \\ \hline
			\rotatebox{90}{$\quad\ \ \ket{-\nicefrac{5}{2}}_D$} & &
			  \fboxrule0pt\fboxsep4pt\fbox{\includegraphics[width=0.2\textwidth]{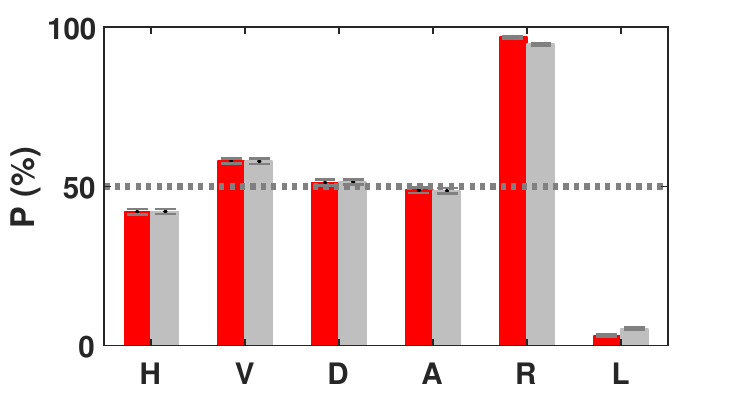}} &  
			  \fboxrule0pt\fboxsep4pt\fbox{\includegraphics[width=0.2\textwidth]{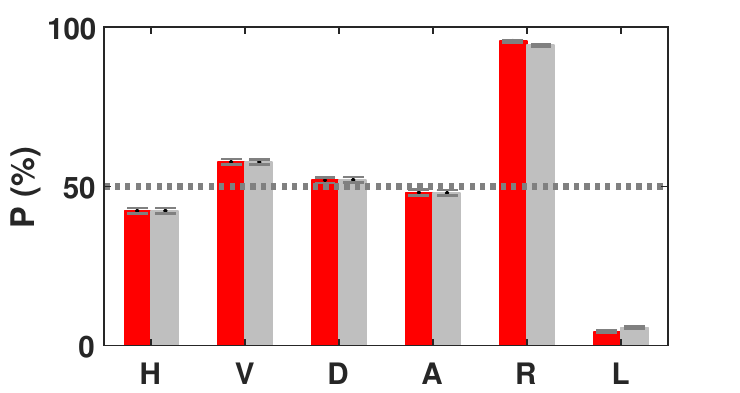}} & 
			  \fboxrule0pt\fboxsep4pt\fbox{\includegraphics[width=0.2\textwidth]{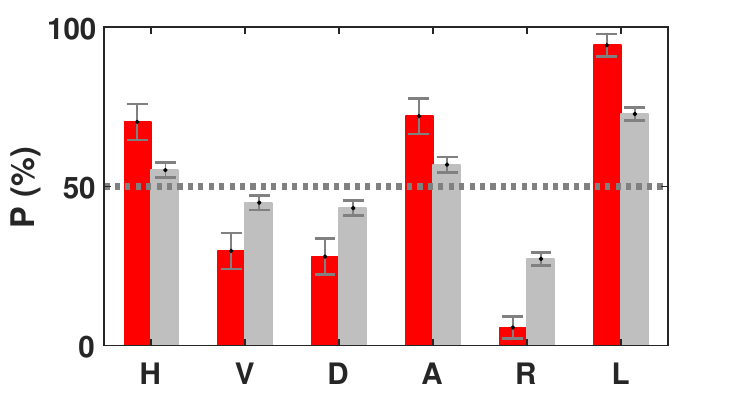}} & 
			  \fboxrule0pt\fboxsep4pt\fbox{\includegraphics[width=0.2\textwidth]{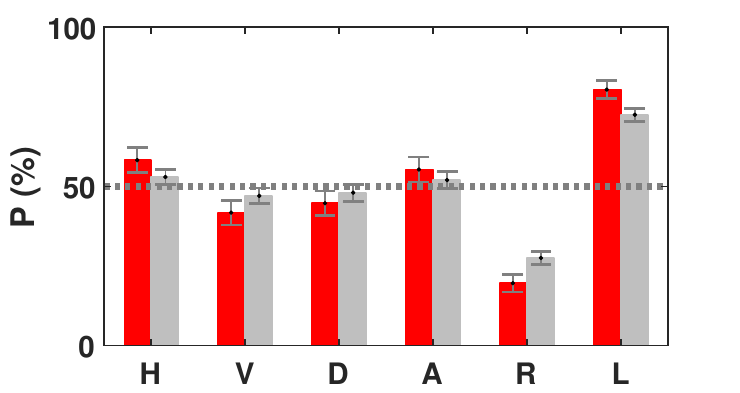}} \\ \hline
			\rotatebox{90}{$\quad\ \ \ket{+\nicefrac{5}{2}}_D$} & &
			  \fboxrule0pt\fboxsep4pt\fbox{\includegraphics[width=0.2\textwidth]{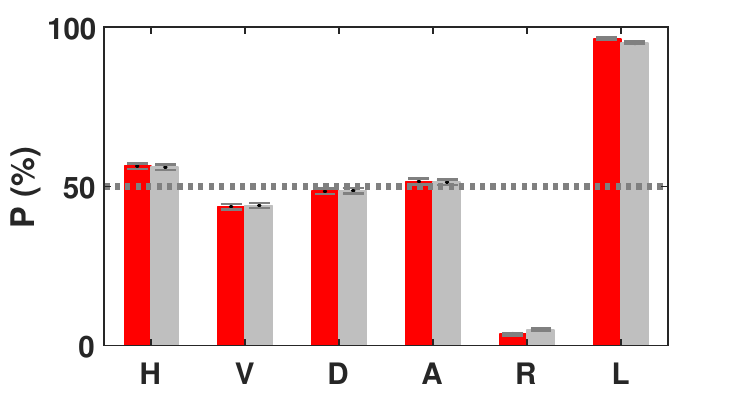}} & 
			  \fboxrule0pt\fboxsep4pt\fbox{\includegraphics[width=0.2\textwidth]{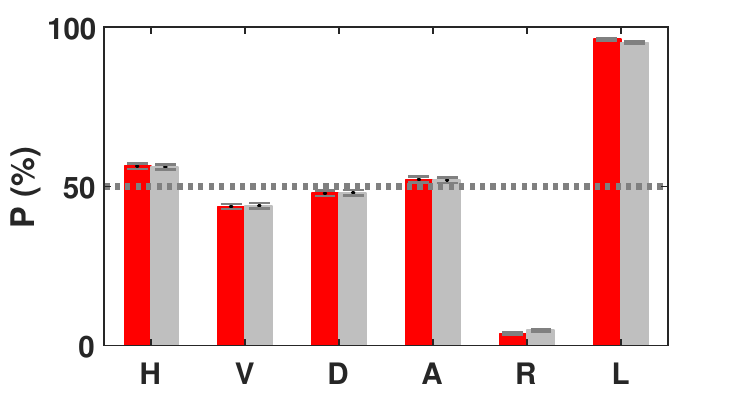}} & 
			  \fboxrule0pt\fboxsep4pt\fbox{\includegraphics[width=0.2\textwidth]{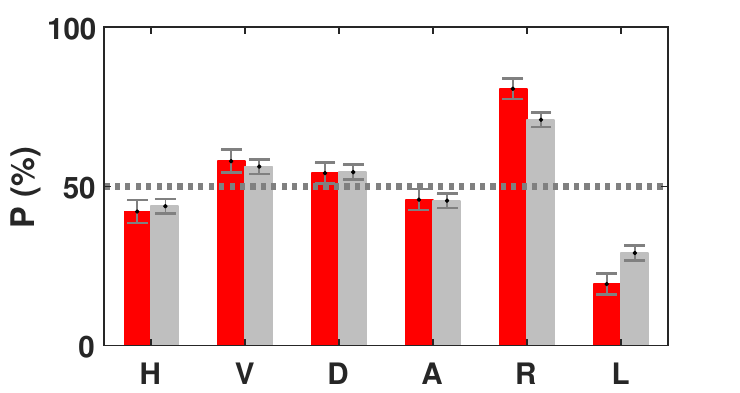}} & 
			  \fboxrule0pt\fboxsep4pt\fbox{\includegraphics[width=0.2\textwidth]{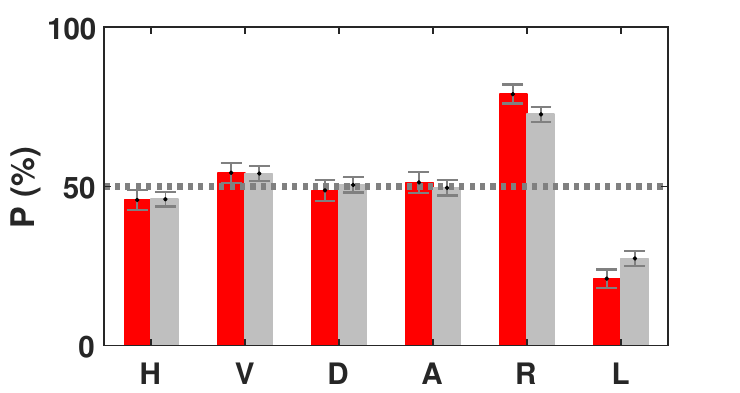}} \\\hline 
		\end{tabular} 
		\caption{\label{fig.TeleportData} Teleportation measurement data evaluated for a exposure time of \SI{350}{\micro\second}. Gray bars and lines show the data without correction. The coloured ones with correction.}
	\end{figure*}

For the teleportation measurement, we calculate histograms of the correlated events of Bell-state measurements and partner-photon polarization. Figure \ref{fig.TeleportData} contains the teleportation measurement data. The columns correspond to the Bell-state-measurement outcome. The first three rows show the probabilities for the superposition input to detect $\ket{H}_{\rph}$, $\ket{D}_{\rph}$, and $\ket{R}_{\rph}$ in dependence of the superposition phase. The probabilities to detect $\ket{V}_{\rph}$, $\ket{A}_{\rph}$, and $\ket{L}_{\rph}$ are derived by calculating $1-\text{P}(\phi)_{(H/D/L)}$. The last two rows contain the probabilities when starting with the energy eigenstates  $\ket{\pm\nicefrac{5}{2}}_D$.

All probabilities are fed into a maximum likelihood algorithm to reconstruct the process matrix.

%\begin{figure*}
	%\mbox{}
%\end{widetext}

\section{Quantum state teleportation: rates and efficiencies}
In total we performed $N_{run} = \num{511670886}$ measurement runs with $\tau_{exp} = \SI{350}{\micro\second}$ exposure time per run and recorded $N_{c,\text{first}} = \num{89838}$ coincident events between the \SI{393}{\nano\meter} heralds and the \SI{854}{\nano\meter} photons for the first passage and $N_{c,\text{second}} = \num{11322}$ for the second passage. The success probabilities per exposure time 
%$\eta_{\text{success,run,first}}$ and $\eta_{\text{success,run,second}}$ 
are then
\begin{align}
    \eta_{\text{success,run,first}} = \frac{N_{c,\text{first}}}{N_{run}} = \num{1.76e-04}\\
    \eta_{\text{success,run,second}} = \frac{N_{c,\text{second}}}{N_{run}} = \num{2.21e-05}
\end{align}
The total exposure time $T_{\text{tot,exposure}}$ was $N_{run} \cdot \tau_{exp} = \SI{1.79e+05}{\second} = \SI{49.75}{\hour}$. With our pair rate per pump power of $\SI{5.17e4}{(\second\, \milli\watt)}^{-1}$ and  $\SI{15}{\milli\watt}$ pump power we get the total number of generated pairs $N_{pairs} = \num{1.388803e+11}$. The success probabilities per pair %$\eta_{\text{success,pair,first}}$ and $\eta_{\text{success,pair,second}}$
are then
\begin{align}
    \eta_{\text{success,pair,first}} = \frac{N_{c,\text{first}}}{N_{pair}} = \num{6.47e-07}\\
    \eta_{\text{success,pair,second}} = \frac{N_{c,\text{second}}}{N_{pair}} = \num{8.15e-08}
\end{align}

These values are composed of several contributions. The detection efficiency of the \SI{854}{\nano\meter} photons in arm B including APD quantum efficiency, spectral filters, fibers and all other optical components is $\eta_{854,B} = \SI{12.6}{\percent}$. The total coupling efficiency from the photon-pair source to the ion setup (in front of the vacuum chamber) is $\eta_{854,A} = \SI{30}{\percent}$. For detecting the 393-nm absorption herald, the HALO covers \SI{4}{\percent} solid angle; in combination with the emission pattern of $\sigma$-transition it collects $\SI{6}{\percent}$ of the emitted \SI{393}{\nano\meter} photons. The \SI{393}{\nano\meter} photon detection efficiency after the HALO is \SI{37.5}{\percent}. The total \SI{393}{\nano\meter} detection efficiency is then $\eta_{393} = \SI{1.64}{\percent}$. We include coincidences between absorption herald and arm B photon within a gate of $\pm \SI{84}{\nano\second}$, corresponding to $\eta_{gate} = \SI{99.9}{\percent}$ of the \SI{854}{\nano\meter} photon wave packet. The remaining part is the absorption efficiency $\eta_{abs}$ per incoming photon. We calculate these probabilities %by dividing $\eta_{\text{success,pair,first}}$ or $\eta_{\text{success,pair,second}}$ by the known efficiencies and get
as
\begin{align}
	\eta_{\text{abs,first}} = \frac{\eta_{\text{success,pair,first}}}{\eta_{854,A} \ \eta_{854,B}\  \eta_{393} \ \eta_{gate}} = \num{1.04e-03}\\
	\eta_{\text{abs,second}} = \frac{\eta_{\text{success,pair,second}}}{\eta_{854,A} \ \eta_{854,B}\  \eta_{393} \ \eta_{gate}} = \num{1.32e-04}
\end{align}
The absorption probability per photon is in good agreement with earlier measured values \cite{Brito2016,Lenhard2015,PhDStephan}.

\section{Quantum state teleportatoin: Error budget \& Improvements}
\begin{figure}
    \centering
    \includegraphics{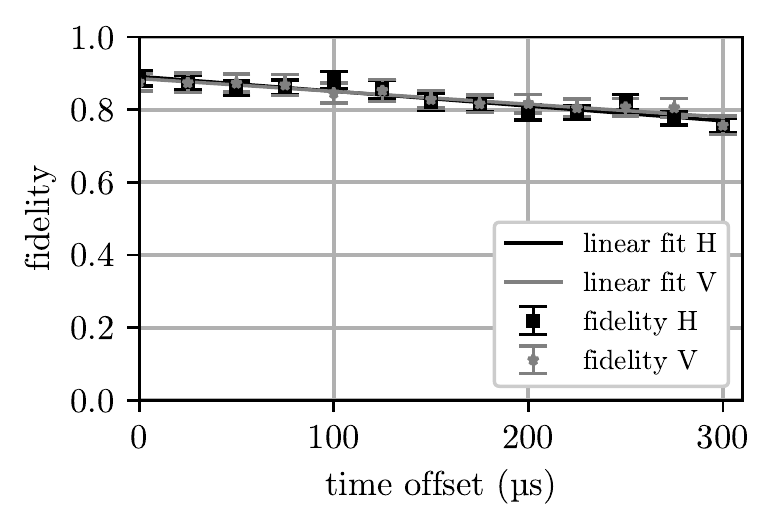}
    \caption{Atom-photon fidelity depending on the time offset for a fixed detection window of $\SI{50}{\micro\second}$ }
    \label{fig:DetectionWindowOffset}
\end{figure}

To investigate the influence of the magnetic field fluctuation, we vary the offset of the detection window (50~µs), taking only later events into account. The data is shown in Fig. \ref{fig:DetectionWindowOffset}. We see a linear decrease in fidelity. Therefore we claim that the fidelity is mainly limited by decoherence due to magnetic field fluctuations.

One improvement, especially for the entanglement transfer, is to avoid the influence of the Larmor phase. To describe the qubit without this phase we need to deal with the different energy spacings of the qubits. This can be done by simultaneously changing the frequency of the reference oscillator to the Larmor frequency of the D-state qubit, while the electron is in the D-state. Once the herald of absorption is detected, the frequency is changed back to the Larmor frequency of the ground-state qubit. In this way, the reference oscillator is always in phase. 

\newpage
\subsection*{Supplementary References}